\documentclass[conference,letterpaper]{IEEEtran}

\usepackage{arydshln}
\usepackage{multirow}
\usepackage[framed,autolinebreaks,useliterate]{mcode}
\usepackage{mathtools}
\usepackage{nicematrix}
\usepackage{float}
\usepackage{tabularx}
\usepackage{booktabs}
\usepackage{cancel}
\usepackage{enumitem} 
\usepackage{ytableau}
\usepackage{graphicx} 
\graphicspath{ {images/} }

\usepackage{color, colortbl}
\usepackage{multirow}

\usepackage{cuted}
\usepackage[linesnumbered,ruled,vlined]{algorithm2e} 

\usepackage[utf8]{inputenc}
\usepackage[english]{babel}
\usepackage{amsthm,amssymb}
\usepackage{amsmath}
\usepackage{bm}
\usepackage{optidef}
\usepackage{mathrsfs}

\usepackage[font=small,labelfont=bf,labelsep=space]{caption}
\usepackage{xcolor}
\def\BibTeX{{\rm B\kern-.05em{\sc i\kern-.025em b}\kern-.08em
    T\kern-.1667em\lower.7ex\hbox{E}\kern-.125emX}}

\usepackage[style=ieee,backend=biber]{biblatex}
\AtBeginBibliography{\footnotesize} 

\usepackage{tikz}
\usetikzlibrary{matrix,backgrounds,positioning}

\usetikzlibrary{calc,patterns,decorations.pathreplacing,decorations.pathmorphing,matrix,positioning,arrows.meta,fit,bending}

\theoremstyle{definition}
\newtheorem{example}{Example}

\newtheorem{lemma}{Lemma}

\newtheorem{definition}{Definition}

\newtheorem{remark}{Remark}

\DeclareMathOperator*{\argmin}{arg\,min}

\DeclareMathOperator{\bin}{bin}

\DeclareMathOperator{\w}{w}
\DeclareMathOperator{\dist}{d}
\DeclareMathOperator{\PW}{PW}

\newcommand{\wm}{\w_{\min}}

\newcommand{\I}{\mathcal{I}}

\renewcommand{\S}{\mathcal{S}}
\newcommand{\Z}{\mathcal{Z}}

\newcommand{\Q}{\mathcal{Q}}

\newcommand{\T}{\mathcal{T}}

\newcommand{\M}{\mathcal{M}}

\newcommand{\C}{\mathcal{C}}

\newcommand{\bc}{\boldsymbol{c}}

\newcommand{\bv}{\boldsymbol{v}}

\newcommand{\bG}{\boldsymbol{G}}

\newcommand{\ind}{\operatorname{ind}}
\newcommand{\ev}{\operatorname{ev}}

\newcommand{\ft}{\mathbb{F}_2}

\newcommand{\Alow}{{\rm LTA}(m,2)}

\newcommand{\Mon}{\mathcal{M}_{m}}
\newcommand{\weako}{\preceq_w}

\IEEEoverridecommandlockouts                              

\title{
Towards Weight Distribution-Aware Polar Codes 
}

\author{
\IEEEauthorblockN{Mohammad Rowshan}
\IEEEauthorblockA{
University of New South Wales\\ 
Sydney, Australia\\ 
Email: m.rowshan@unsw.edu.au \vspace{-0.5em}}
\and
\IEEEauthorblockN{Vlad-Florin Dr\u{a}goi}
\IEEEauthorblockA{
Aurel Vlaicu University\\
Arad, Romania\\
Email: vlad.dragoi@uav.ro \vspace{-0.5em}}

}

\begin{document}

\maketitle
\pagestyle{plain}
\pagestyle{empty}

\begin{abstract}
Polar codes are constructed based on the reliability of sub-channels resulting from the polarization effect. However, this information-theoretic construction approach leads to a poor weight distribution. To address this issue, pre-transformed polar codes, such as CRC-polar codes and PAC codes, have been employed. In this paper, we focus on the structure of polar codes without applying any pre-transformations and explore methods, guided by the weight-contribution partial order, to design polar-like codes with enhanced weight distribution, notably without employing any search or optimization algorithms. Numerical results demonstrate improvement over a range of codes both with and without pre-transformation. 
\end{abstract}



\section{Introduction}
Polar codes 
\cite{arikan} are members of a larger family of codes called decreasing monomial codes \cite{bardet2016crypt,dragoi17thesis}, which also includes Reed--Muller codes. 
Building on classical results related to Reed--Muller codes \cite{kasami1970weight}, recent studies \cite{bardet,vlad1.5d,ye2024distribution,rowshan2024weight} have delved into the algebraic properties of polar codes and, in general, decreasing monomial codes, among which a subgroup of their permutation group called the lower triangular affine group ($\Alow$). Furthermore, they shed light on the weight structure, offering closed-form formulas for the enumeration of codewords with a weight up to twice the minimum weight ($\wm$). 
The permutation group has a significant contribution in ensemble decoding \cite{GEECB21,  PBL21, IU22}. 
Furthermore, the formation of minimum weight codewords of polar and PAC codes was studied in terms of the sum of rows in the polar transform in \cite{rowshan2023formation} and  algorithmic methods for the enumeration of PAC codes were investigated in \cite{rowshan2023minimum,ellouze2023low,zunker2024enumeration,ellouze2024computing}. The insights from these works were used in \cite{gu2023rate,gu2024reverse,gu2025pac} for code design and analysis.

Polar codes are constructed by selecting the most reliable sub-channels of the polarized vector channel. The reliability of sub-channels is evaluated using various methods, such as Bhattacharyya parameter \cite{arikan} for binary erasure channels (BEC), high-precision density evolution (DE) \cite{mori},  Tal-Vardy method \cite{tal2} and the Gaussian approximation (GA) method \cite{chung,trifonov}, both with lower complexity. There are also low complexity SNR-independent methods for reliability evaluation, such as partial ordering \cite{bardet,schurch} and polarization weight \cite{he}.


In this study, we analyze the information set that defines the codes in terms of both reliability and weight contribution with respect to $m$ and discover some associated properties. Then, we propose approaches to strike a balance between the reliability and weight distribution of a code, resulting in significant improvement in performance. 

\section{Polar codes as Decreasing monomial codes}%
We denote by $\ft$ the finite field with two elements. 
Also, subsets of consecutive integers are denoted by $[\ell,u]\triangleq\{\ell,\ell+1,\ldots,u\}$. 
The cardinality of a set is denoted by $|\cdot|$ and the set difference by $\backslash$. 
The Hamming \emph{weight} of a vector $\bc \in \ft^N$ is $\w(\bc)$. 
The binary numbers $\bin(i)=(i_{m-1}\dots i_1i_0)_2$ corresponding to $(i_0,\dots,i_{m-1})
\in \ft^m$, where $i=\sum_{j=0}^{m-1} i_j2^j$, are written so that $i_{m-1}$ is the most significant bit. 
A $K$-dimensional subspace $\C$ of $\ft^N$ is called a linear $(N,K,d)$ \emph{code} over $\ft$ 
where $d$ is the minimum distance of code $\C$; that is,
$\dist(\C) \triangleq \min_{\bc,\bc' \in \C, \bc \neq \bc'} \dist(\bc,\bc')=\wm$, which is equal to the minimum weight of codewords excluding the all-zero codeword.
Let us collect all $\w$-weight codewords of $\C$ in set $W_{\w}$ as
$
    W_{\w}(\C) = \{ \bc\in\C \mid \w(\bc)=\w \}.
$
A \emph{generator matrix} $\bG$ of an $(N,K)$-code $\C$ is a $K \times N$ matrix in $\ft^{K\times N}$ whose rows are $\ft$-linearly independent codewords of $\C$. Then $\C = \{\bv \bG \colon \bv \in \ft^K\}$.

\subsection{Monomials and monomial codes}
Let $m$ be a fixed integer that represents the number of different variables 
and $m$-variable monomial  $\mathbf{x}^{\bin(i)}=x_0^{i_0}\cdots x_{m-1}^{i_{m-1}}.$
Denote the set of all monomials by 
$
\Mon\!\triangleq\!\left\{\mathbf{x}^{\bin(i)} \mid\bin(i) \in \mathbb{F}_{2}^{m}\right\}
.$
For example, for $m=2$ we have $\Mon=\{x_0x_1,x_1,x_0,\bm{1}\}.$ Furthermore, the \emph{support of monomial} $f=x_{l_1}\dots x_{l_s}$ is $\ind(f)=\{l_1,\dots,l_s\}$ and its degree is $\deg(f)=|\ind(f)|.$ The degree induces a ranking on any monomial set $\I\subseteq\Mon$, that is, $\I=\bigcup_{j=0}^{m}\I_j$, where $\I_j=\{f\in\I\mid\deg(f)=j\}.$
\begin{remark}
    By convention, we use the indices of the most reliable sub-channels to define polar codes. These indices correspond to $\overline{\bin(i)}$ of all monomials $\mathbf{x}^{\bin(i)}\in\I$. For example, the index $12=(1100)_2$ for $m=4$ corresponds to $x_0x_1$ where $\bin(i)=(0011)_2$. 
    Throughout this paper, we may use sub-channel indices and monomials, or decimal and binary representation of indices interchangeably. 
\end{remark}



\begin{definition}[Monomial code]
Let $\I\subseteq\Mon.$ A monomial code defined by $\I$ is the vector subspace $\C(\I) \subseteq \ft^{2^m}$ 
generated by $\{ \ev(f) ~|~ f \in \I\}$, where the evaluation function $\ev(f)$ associates $f$ with a binary vector in $\ft^{2^m}$. 
\end{definition}



Let ``$|$" denote the divisibility between monomials, i.e., $f|g$ iff $\ind(f)\subseteq\ind(g).$ Also, the greatest common divisor of two monomials is $\gcd(f,g)=h$ with $\ind(h)=\ind(f)\cap\ind(g).$

\begin{definition}\label{def:order}Let $m$ be a positive integer and $f,g\in\Mon.$ Then $f\weako g$ if and only if $ f|g.$ When $\deg(f)=\deg(g)=s$ we say that $f\preceq_{sh} g$ if $\forall\;1\leq\ell\leq s\;\text{ we have }\;  i_\ell \le j_\ell$, where $f=x_{i_1}\dots x_{i_s}$, $g=x_{j_1}\dots x_{j_s}$. 
Define $f\preceq g\quad \text{iff}\quad \exists g^*\in \Mon\;\text{s.t.}\; f\preceq_{sh} g^*\weako g$.
\end{definition}
These relations define partial orders for the reliability of channels corresponding to monomials in $\I$.  \cite{mori,bardet,schurch,Wang-Dragoi2023,
Wu-Siegel2019,Dragoi-CristescuPO2021}. 
\begin{definition}\label{def:dec_set}
      A set $\I \subseteq \Mon$ is \emph{decreasing}  if and only if ($f \in \I$ and $g \preceq f $) implies $g \in \I$. 
\end{definition}

Any monomial code $\C(\I)$ with $\I$ decreasing is called \emph{decreasing monomial code}. Polar and Reed--Muller codes are decreasing monomial codes \cite{bardet}. 


\subsection{Small weight codewords and their significance}\label{ssec:enum}
Let $\I$ be a decreasing monomial set and $r=\max_{f\in \I}\deg(f).$ Then, the number of codewords with weights  $\wm=2^{m-r}$ and $1.5\wm$ of any decreasing monomial codes including polar codes are determined by \cite{bardet,vlad1.5d}
\begin{equation}\label{eq:sum_A_wm}
    |W_{\wm}(\I)|=\sum\limits_{f\in \I_r}2^{r+|\lambda_f(f)|},
\end{equation}
\begin{equation}\label{eq:formula_15w}          |W_{1.5\wm}|=\smashoperator{\sum\limits_{\substack{f,g\in \I_r\\ h=\gcd(f,g)\in\I_{r-2}}}} \frac{2^{r+2+|\lambda_h|+|\lambda_{f}(\frac{f}{h})|+|\lambda_{g}(\frac{g}{h})|}}{2^{\alpha_{\frac{f}{h},\frac{g}{h}}}},
\end{equation} 
where 
\begin{equation}
    |\lambda_f(g)|=\sum_{i\in\ind(g)}\lambda_f(x_i),
\end{equation}
for $\lambda_f(x_i)=|\{j\in[0,i) \mid j\notin\ind(f)\}|$ which is the degree of freedom we have on $x_i, i\in\ind(g)$ and $\alpha_{{\frac{f}{h}},{\frac{g}{h}}}$ is the degree of collision defined in \cite[Def. 8]{vlad1.5d}. 
The closed-form formula for other weights up to $2\wm$ can be found in \cite{ye2024distribution,rowshan2024weight}. 
\begin{remark}\label{rem:lambda_contrib}
    Since the parameter $|\lambda_f(f)|$ or, in general, $|\lambda_f(g)|$ appears in \eqref{eq:sum_A_wm} and \eqref{eq:formula_15w}, any code that yields smaller values for these parameters would exhibit a better weight distribution. 
\end{remark}
The block error rate (BLER) of a code $\C$ under soft-input maximum likelihood (ML) decoding over binary-input discrete memoryless channel (BI-DMC) is bounded by the well-known \emph{union-Bhattacharyya bound} \cite{stark2023introduction}: 
\begin{equation}\label{eq:union_Bh_bound}
    P_{e}\leq \sum_{\mathclap{\w=\wm}}^{N} |W_{\w}(\C)| \gamma^{\w},
\end{equation}
where $\gamma=\mathrm{exp}(-R\cdot E_{\mathrm{b}}/N_0)$ is the Bhattacharyya parameter for the AWGN channel,  $E_b/N_0$ is the normalised SNR per information bit and $R$ is the code rate. Observe that $\gamma^{\w}$ decreases exponentially by $\w$ and the corresponding terms in \eqref{eq:union_Bh_bound} can be discarded; Hence, $|W_{\w}(\C)|$ for small $\w$ plays significant role in obtaining smaller BLER.

\subsection{Performance of $\C(\I)$ under SC decoding}\label{ssec:SC_perf}
If we decode a polar code formed by set $\I$ with successive cancellation (SC) decoding algorithm, the block error event $E$ is a union over $\I$ of the event that the first bit error occurs, denoted by $E_i \triangleq \{\hat{u}_i \neq u_i \mid \hat{u}_1^{i-1}=u_1^{i-1}\}$ where $E=\bigcup_{i\in\I} E_i$. Let $E^c$ denote the event that a received sequence associated to a codeword is correctly decoded, that is, $\hat{u}_1^N=u_1^N$, then the probability of block error is obtained by \cite{mori}
\begin{equation}\label{eq:SC_BLER}
    P_{SC}(\I) = P(E)=1-P(E^c)=1-\prod_{i \in \I}(1-P(E_i)),
\end{equation}
where $P(E_i)$ is the probability of error at sub-channel $i$ at a particular noise power or SNR given that bits 0 to $i-1$ are decoded successfully. Note that $P(E_i)=0$ for any $i\in\I^c$ and (roughly) inversely proportional to any reliability metric such as mean log-likelihood ratio (LLR), Bhattacharyya parameter, or polarization weight (PW) of the corresponding sub-channel with monomials in $\I$. Thus, weight distribution alone does not fully account for the performance of polar codes. The reliability of the sub-channels associated with the monomials must also be considered.

\section{Reliability Distribution of monomials}\label{sec:pw_prop}

As an initial step toward balancing both reliability and weight distribution for code design, we analyze the distribution of the sub-channel reliability associated to the monomials of the same degree, as well as monomials with degree differences of 1. This analysis linked with the analysis of weight contribution of monomials in the next section enables us to adjust and the weight distribution of polar codes. 

To estimate reliability, we adopt the polarization weight or beta expansion metric from \cite{he}, which is defined as 
\begin{equation}\label{eq:pw}
    \PW(i)=\sum_{j=0}^{m-1} i_j \cdot \beta^j,
\end{equation}
where $i$ is the index of sub-channel, $\beta>0$. 
Here, we assume $\beta=2^{\frac{1}{4}}$.   
This reliability metric offers a computationally tractable tool for our analysis, although it lacks accuracy. 
\subsection{Reliability distribution of same-degree monomials}
Each bit position $j$ contributes a term $i_j \cdot \beta^j$ to $\PW(i)$. 
Higher bit positions (closer to $m-1$ ) contribute larger values due to the exponential increase of $\beta^j$ when $\beta>1$. 

\begin{example}\label{ex:WP_max-min}
For $m=8$, $\beta>0$, and Hamming weight $t=4$, the maximum $\PW(i)$ occurs when the 1's occupy the highest bit positions; $\underbrace{\tiny 11\cdots11}_{(t-1)\times \text{`1'}}\underbrace{\tiny 00\cdots00}_{(m-t+1)\times \text{`0'}}$ (here $i=11110000_2$):
$$
\PW(i)_{\max }=\beta^4+\beta^5+\beta^6+\beta^7 \approx 10.57.
$$
The minimum $\PW(i)$ occurs when the 1's occupy the lowest bit positions; $\underbrace{00\cdots00}_{(m-t)\times \text{`0'}}\underbrace{11\cdots11}_{(t)\times \text{`1'}}$ (here, $i=$ $00001111_2$ ):
$$
\PW(i)_{\min }=\beta^0+\beta^1+\beta^2+\beta^3 \approx 5.285.
$$
Since the powers of $\beta$ grow exponentially, the distribution of $\PW(i)$ values in $\left[5.285, 10.57\right]$ will not be uniform, but will spread more widely (the difference between adjacent $\PW(i)$'s increases) toward $\PW(i)_{\min}$ and $\PW(i)_{\max}$.
\end{example}

Similarly, increasing the Hamming weight from $t-1$ to $t$ for a given $m$ involves adding one additional bit set to 1 at a particular position, which increases $\PW(i)$ by $\beta^j$, resulting in a wider spread of $\PW(i)$. 
\subsection{Reliability difference b/w monomials of different degrees}
Considering Hamming weights larger than $m/2$ (or alternatively smaller than $m/2$), we compare the distribution of polarization weights between indices with Hamming weights $t$ and $t-1$ (or alternatively $t$ and $t+1$). 

\begin{figure}
    \centering
    \includegraphics[width=0.75\linewidth]{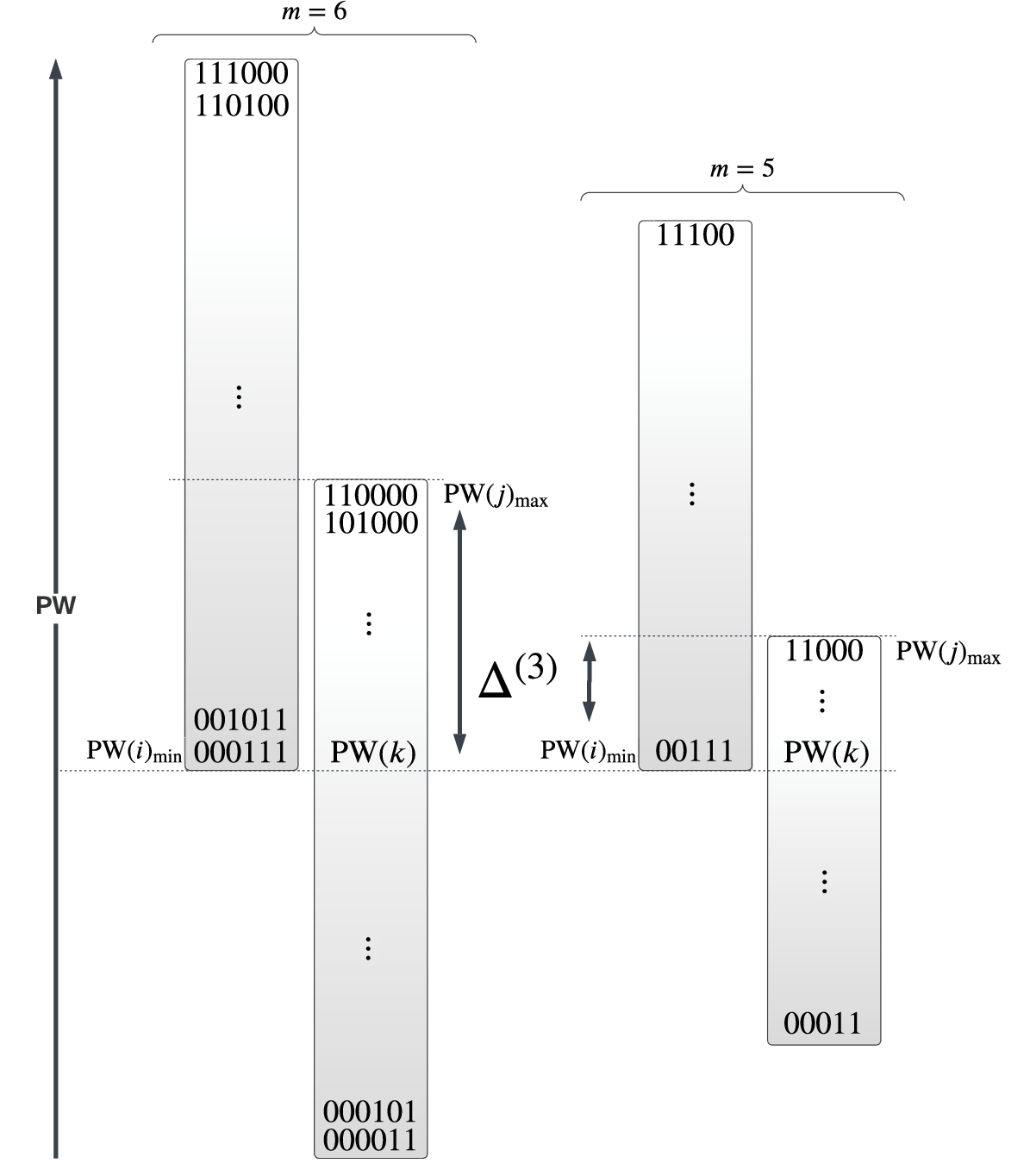}
    \caption{An example demonstrating the effect of increase in $m$ on the overlap of sub-channels with different weights.}
    \label{fig:ex_pw_ovrlap}
\end{figure}

Defining $\S_t=\{i\!\in\![0,N\!-\!1]\mid \w(i)\!=\!t\}$, the difference between $\PW(i)_{\min}$ and $\PW(j)_{\max}$ for a specific $t$ is  
\begin{equation}\label{eq:delta}
\Delta^{(t)} = \PW^{(t-1)}_{\max} - \PW^{(t)}_{\min}.
\end{equation}
As $m$ increases, $\mathrm{PW}(j)_{\max }$ grows much faster than $\mathrm{PW}(i)_{\min }$ due to the dominance of $\beta^{m-1}$, which far outweighs the contribution of smaller powers of $\beta$ in $\mathrm{PW}(i)_{\min }$, although it increases with the addition of one 1 in its binary representation. This will result in a significant growth of $\Delta^{(t)}$ of the $\PW(i)$ distributions for the adjacent Hamming weights $t$ and $t-1$, as $t$ reduces in direction $m\rightarrow m/2$. 
Fig. \ref{fig:ex_pw_ovrlap} illustrates the increase in overlap, where the index $k$ is obtained by 
\[
k = \argmin_{k:\w(k)=\w(j)} \left|\PW(k)-\PW(i)_{\min}\right|,
\]
which is the closest sub-channel $j$ with Hamming weight $\w(j)=t-1$ to the least reliable sub-channel $i$ with larger Hamming weight $\w(i)=t$.

\subsection{Symmetry in $\PW(i)$ with complementary weights}
For $\bin(i)$ with $\w(i)=t$, consider its complement, 
$$
\overline{\bin(i)}=\left(\left(1-i_{m-1}\right)\left(1-i_{m-2}\right) \cdots\left(1-i_0\right)\right)_2, 
$$
where the Hamming weight is $m-t$, since $\bar{i}_j=1-i_j$ flips all bits of $\bin(i)$. Then, the polarization weight of $\overline{\bin(i)}$ is
$$
\PW(\bar{i})=\sum_{j=0}^{m-1} \bar{i}_j \cdot \beta^j=\sum_{j=0}^{m-1}\left(1-i_j\right) \cdot \beta^j = \sum_{j=0}^{m-1} \beta^j-\PW(i).
$$
Let $S_m=\sum_{j=0}^{m-1} \beta^j=\frac{\beta^m-1}{\beta-1}$, which is a geometric series sum if $\beta \neq 1$. Then, the polarization weights $\PW(i)$ and $\PW(\bar{i})$ are symmetric around $S_m / 2$; that is,
\begin{equation}
    \PW(\bar{i})=S_m - \PW(i).
\end{equation}
Furthermore, for every $m$, the number of all possible indices with Hamming weight $t$ and $m-t$ is equal as 
$$
    {m \choose t}={m \choose m-t}. 
$$
Therefore, the distributions of $\PW(i)$ for the indices of the Hamming weights $t$ and $m-t$ are reflections of each other around $S_m / 2$. 



\section{Weight Contribution of Monomials} \label{sec:wt_prop} 
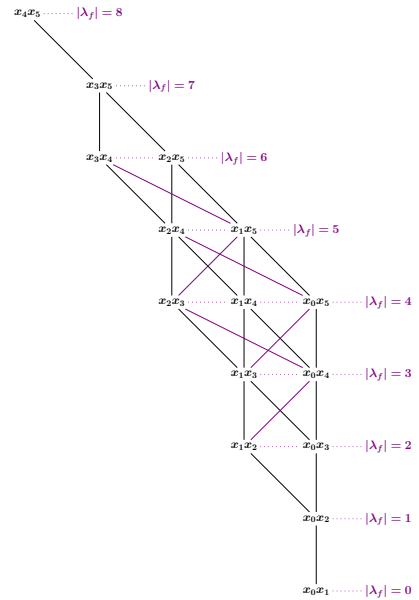
\begin{figure}[!ht]
\centering\resizebox{0.3\textwidth}{!}{
\begin{tikzpicture}[scale=3,thick]

\node at (0,0) (01) {\Large$\bm{x_0x_1}$};
\node at (1,0) (l1) {\Large$\color{violet} \bm{|\lambda_f|= 0}$};

\node at (0,1) (02) {\Large$\bm{x_0x_2}$};
\node at (1,1) (l2) {\Large$\color{violet}\bm{|\lambda_f|= 1}$};

\node at (0,2) (03) {\Large$\bm{x_0x_3}$};
\node at (1,2) (l3) {\Large$\color{violet}\bm{|\lambda_f|= 2}$};

\node at (0,3) (04) {\Large$\bm{x_0x_4}$};
\node at (1,3) (l4) {\Large$\color{violet}\bm{|\lambda_f|= 3}$};

\node at (0,4) (05) {\Large$\bm{x_0x_5}$};
\node at (1,4) (l5) {\Large$\color{violet}\bm{|\lambda_f|= 4}$};

\node at (-1,2) (12) {\Large$\bm{x_1x_2}$};
\node at (-1,3) (13) {\Large$\bm{x_1x_3}$};
\node at (-1,4) (14) {\Large$\bm{x_1x_4}$};
\node at (-1,5) (15) {\Large$\bm{x_1x_5}$};
\node at (0,5) (l15) {\Large$\color{violet}\bm{|\lambda_f|= 5}$};

\node at (-2,4) (23) {\Large$\bm{x_2x_3}$};
\node at (-2,5) (24) {\Large$\bm{x_2x_4}$};
\node at (-2,6) (25) {\Large$\bm{x_2x_5}$};
\node at (-1,6) (l25) {\Large$\color{violet}\bm{|\lambda_f|= 6}$};

\node at (-3,6) (34) {\Large$\bm{x_3x_4}$};
\node at (-3,7) (35) {\Large$\bm{x_3x_5}$};
\node at (-2,7) (l35) {\Large$\color{violet}\bm{|\lambda_f|= 7}$};

\node at (-4,8) (45) {\Large$\bm{x_4x_5}$};
\node at (-3,8) (l45) {\Large$\color{violet}\bm{|\lambda_f|= 8}$};

\draw[line width=0.25mm](01) --(02) --(03) -- (04) --(05);
\draw[line width=0.25mm](02) --(12) --(13) -- (14) -- (15);
\draw[line width=0.25mm](03) --(13) --(23) -- (24) -- (25);
\draw[line width=0.25mm](04) --(14) --(24) -- (34) -- (35);
\draw[line width=0.25mm](05) --(15) --(25) -- (35) -- (45);

\draw[line width=0.25mm,violet](12) --(04) --(23) -- (15) -- (34);
\draw[line width=0.25mm,violet](13) --(05) --(24) ;

\draw[line width=0.25mm, loosely dotted,  violet]  (01) -- (l1);
\draw[line width=0.25mm, loosely dotted,  violet]  (02) -- (l2);
\draw[line width=0.25mm, loosely dotted,  violet]  (12) -- (03) -- (l3);
\draw[line width=0.25mm, loosely dotted,  violet]  (13) -- (04) -- (l4);
\draw[line width=0.25mm, loosely dotted,  violet]  (23) -- (14) -- (05) -- (l5);
\draw[line width=0.25mm, loosely dotted,  violet]  (24) -- (15) -- (l15);
\draw[line width=0.25mm, loosely dotted,  violet]  (34) -- (25) -- (l25);
\draw[line width=0.25mm, loosely dotted,  violet]  (35) -- (l35);
\draw[line width=0.25mm, loosely dotted,  violet]  (45) -- (l45);



\end{tikzpicture}
}
\caption{{\bf Contribution diagram}: Monomials of degree $s=2$ for $m=6$, partially ordered in terms of reliability (black lines) and weight contribution (both black and violet lines). Monomials with identical $|\lambda_f|$ contributes the same number of codewords with weight $2^{6-2}$.}
\label{fig:monomials_deg2m6}
\end{figure}

In \cite{rowshan2024weight}, we proposed a new partial order between monomials of equal degree that indicates the relationship between the contributions of monomials to the weight distribution. 
\begin{definition}
The order relation between $f,g$ of equal degree $s=m-t$ based on the weight contribution is
\[\forall\;f,g\in \I_s,\text{ we say }\;f\leq_{\wm} g \text{ if }|\lambda_f|\leq |\lambda_g|.\]
\end{definition}
According to Remark \ref{rem:lambda_contrib}, this means that monomial $f$ contributes less than $g$ to the total number of minimum weight codewords $|W_{\wm}|$ since $|\lambda_f|\leq |\lambda_g|$. 
Typically, we divide the set of all monomials into disjoint subsets, first according to their degree (the $s^{th}$ subset corresponds to $\I_s$) as
\begin{equation}\label{eq:subsets}
    \Z_s=\{f\in\M_m\mid \deg(f)=s\},
\end{equation}
and then with respect to $\leq_{\wm}$. Fig. \ref{fig:monomials_deg2m6} shows an example of such a subset for all monomials of degree $s=2$.

Discovering more properties induced by $\lambda_f$ becomes necessary to understand the weight contribution of monomials. Similar symmetry relation as the one between $\PW(\bar{i})$ and $\PW({i})$ exists for $\lambda_f$ and $\lambda_{\check{f}}.$ First, notice that$|\lambda_f|\in\{0,s\times (m-s)\}$ for any $f\in\T_s.$ 
\begin{lemma}\label{lma:lambda_bounds}
    For any $f\in\Mon$ with $\deg(f)=s$, we have 
    \begin{equation}
        0 \leq |\lambda_f| \leq s(m-s).
    \end{equation}
\end{lemma}
Let us give an example to illustrate the symmetry property. 

\begin{example}\label{ex:young-diagrams}
Let $m=7,s=3$ and $f=x_0x_3x_5.$ Then $\check{f}=x_1x_2x_4x_6$ and the partition associated to $f$ and $\check{f}$ is 
 $\lambda_f=(5-2,3-1,0-0)=(3,2,0)$, and $\lambda_{\check{f}}=(6-3,4-2,2-1,1-0)=(3,2,1,1)$, respectively.  The Young diagram of $f$ lies inside the $3\times 4$ grid, while the Young diagram of $\check{f}$ lies inside the $4\times 3$ grid and both are given below.
    \begin{center}\resizebox{0.25\textwidth}{!}{
    \begin{tikzpicture}[inner sep=0in,outer sep=5in]
    \hspace{-100pt}
      \node(m) {
\begin{ytableau}
 *(blue! 40) * &   &  \\
 *(blue! 40) *  &  &  \\
*(blue! 40) *  &*(blue! 40) *   &  \\
*(blue! 40) *  &*(blue! 40) *   &*(blue! 40) *  \\
\end{ytableau}};
\hspace{100pt}
\node(n) {
\begin{ytableau}
  ~~  &   &  &\\
 *(blue! 40) *  & *(blue! 40) * &  &\\
*(blue! 40) *  &*(blue! 40) *   &*(blue! 40) *  &\\
\end{ytableau}};

\node at (-0.75,1) {$x_1$};
\node at (-.25,1) {$x_2$};
\node at (0.25,1) {$x_4$};
\node at (0.75,1) {$x_6$};
\node at (-1.4,0.5) {$x_0$};
\node at (-1.4,0) {$x_3$};
\node at (-1.4,-0.5) {$x_5$};
\node at (0,-1.25) {$\lambda_f$};

\node at (-4,1.25) {$x_0$};
\node at (-3.5,1.25) {$x_3$};
\node at (-3,1.25) {$x_5$};

\node at (-4.6,0.75) {$x_1$};
\node at (-4.6,0.25) {$x_2$};
\node at (-4.6,-.20) {$x_4$};
\node at (-4.6,-0.75) {$x_6$};
\node at (-3.6,-1.5) {$\lambda_{\check{f}}$};

\end{tikzpicture}
}
 \end{center}
Notice that the total number of blue subsets ($\lambda_f$ and $\lambda_{\check{f}}$ together) equals $3\times 4=12.$ 
\end{example}

This example gives an image of the complementary property:
\begin{equation}\label{eq:lambda-f-check}
    |\lambda_f|+|\lambda_{\check{f}}|=s\times (m-s).
\end{equation}

The alternative proof of this result is a direct application of the definition of $|\lambda_f|=\sum_{i\in\ind(f)}i-s(s-1)/2.$ 


Equation \eqref{eq:lambda-f-check} has an implication similar to that of reliability using polarization weight. Indeed, we have that 
\begin{equation}
    f\leq_{\wm}g\Leftrightarrow \check{g}\leq_{\wm} \check{f}.
\end{equation}

This means that we do have a symmetry relative to the middle subset. More exactly, when $m$ is odd, there are two middle subsets $\Z_{{\frac{m-1}{2}}}$ and $\Z_{\frac{m-1}{2}}+1$, while when $m$ is even, we have a single middle subset $\Z_{{\frac{m}{2}}}$. Each subset $\Z_{m-s}$ is a reflection of the subset $\Z_{s}.$

The following lemma establishes a relation between the weight contribution and the reliability of monomials in the subset $\Z_s$. 
\begin{lemma}\label{lem:lambda_reli}
    Let $f,g\in\Mon$ and $f\preceq g.$ Then, $|\lambda_f|<|\lambda_g|$. 
\end{lemma}


Regarding the converse implication, $|\lambda_f|<|\lambda_g|$ does not necessarily imply $f\preceq g.$ Take for example $g=x_9x_2x_1x_0$ and $f=x_5x_4x_3x_1.$ We have $|\lambda_f|=7,|\lambda_g|=6$, however $f,g$ are not comparable w.r.t. to $\preceq.$


In any code design or modification, we need a data structure for $\Z_s, 0\leq s\leq m$ with levels based on $|\lambda_f|$. The property in the following lemma assists in forming the levels in Fig. \ref{fig:monomials_deg2m6}.

\begin{lemma}\label{lma:lambda+1}
    Let $f=x_{i_{s-1}}\cdots x_{i_1}x_{i_0}$ for $i_{s-1}>\cdots>i_1>i_0$ with $f\in\Mon$, degree $\deg(f)=s$, and $|\lambda_f|<s(m-s)$. Then, we can find $g\in\Mon$ of the same degree with 
    $$|\lambda_g|>|\lambda_f|,$$
    by changing any variable $x_{j_1}, j_1\in\mathcal{J}_1=\{i_{s-1},\dots,i_1,i_0\}$ of $f$ to $x_{j_2}$ where $j_2\in\mathcal{J}_2=[j_1+1,m-1]\backslash\mathcal{J}_1$. As a special case, we can get 
    $$|\lambda_g|=|\lambda_f|+1,$$
    by replacing $x_{j_1}$ with $x_{j_1+1}$ given $j_1+1\not\in\ind(f)$. 
\end{lemma}


Algorithm \ref{alg:groups} generates $m+1$ subsets 
of sub-channel indices of identical Hamming weight $m-s$ (for their binary representation), 
where each subset is divided into levels in $\mathbf{L}$ corresponding to their $|\lambda_f|=d,\deg(f)=s$. For each subset, we generate the index of the most reliable sub-channel in the form $(1..10..0)_2$ in line 5. Then, all indices at level $d$ are obtained by swapping the bits of indices at level $d-1$ in line 14. The index $-1$ in $\Z[s][-1]$ indicates the elements of level $d-1$ (last generated level) in subset $\Z[s]$.

\begin{algorithm}[t]
    \footnotesize
    \caption{Generating contribution structure} 
    \label{alg:groups} 
    \DontPrintSemicolon

    \SetKwFunction{FSub}{genZ} 
    \SetKwProg{Fn}{function}{:}{} 
    \SetKwFunction{Unique}{Unique}
    \Fn{\FSub{$m$}}{
        $\Z \gets \{\}$ \tcp*{set of groups}
        \For{$s\gets 0$ \KwTo $m$}{
            Initialize $\Z[s] \gets []$ \tcp*{for degree-$s$ monomials}
            $i \gets \operatorname{bin2dec}([1]\times(m-s)+[0]\times s)$\;
            Append $i$ to $\Z[s]$\;
            \For{$d \gets 0$ \KwTo $s\times(m-s)$}{
                $\mathcal{L} \gets []$ \tcp*{for $|\lambda_f|=d$, Lemma \ref{lma:lambda_bounds}}
                \For{$i \gets \Z[s][-1]$}{
                    $f \gets \bin(i, m)$\;
                    \ForEach{\( j \) \textbf{in} indices of \( f \)}{
                        \If{\( f[j] = 0 \) \textbf{and} \( j > 0 \) \textbf{and} \( f[j-1] = 1 \)}{
                            Create a copy \( g \gets f \)\;
                            Swap \( g[j] \) and \( g[j-1] \): \( g[j] \gets 1 \), \( g[j-1] \gets 0 \) \tcp*{Lemma \ref{lma:lambda+1}}
                            Append \( \operatorname{bin2dec}(g) \) to \( \mathcal{L} \)\;
                        }
                    }
                }
                Append \( \Unique(\mathcal{L}) \) to \( \Z[s] \)\;
            }
        } 
        \KwRet $\Z$;
    }
\end{algorithm}

\section{Code Design: Balancing Reliability and Weight Distribution}\label{sec:code_design}
In this section, based on the analysis in Sections \ref{sec:pw_prop} and \ref{sec:wt_prop}, we propose an approach that adjusts the weight distribution of polar codes without significantly sacrificing reliability aiming to strike a balance between them. 

\begin{table}[]
\caption{The distance properties of codes in Figs. \ref{fig:BLER_128} and \ref{fig:BLER_256}.}
\label{tab:wd}
\centering\footnotesize
\begin{tabular}{|l|l|l|l|l|l|}
\hline
$N$                & $K$ & Scheme & Profile & $d_{\min}$ & $A_{d_{\min}}$ \\ \hline
\multirow{9}{*}{128} &  $24+8$             & CRC-Polar & 5G & 32 & 178 \\ \cdashline{2-2} \cline{3-6} 
                     &  \multirow{2}{*}{24} & PAC    & 5G  & 16 & 8 \\ \cline{3-6} 
                     &                     & PAC      & 5G-WD & 32 & 332 \\ \cline{2-6}  
                     &  $64+8$             & CRC-Polar & 5G & 8 & 8 \\ \cdashline{2-2} \cline{3-6} 
                     & \multirow{2}{*}{64} & PAC      & 5G & 8 & 256 \\ \cline{3-6} 
                     &                     & PAC      & 5G-WD & 16 & 2529 \\ \cline{2-6} 
                     &  $96+8$             & CRC-Polar & 5G & 6 & 20 \\ \cdashline{2-2} \cline{3-6} 
                     & \multirow{2}{*}{96} & PAC      & 5G & 4 & 96 \\ \cline{3-6} 
                     &                     & PAC      & 5G-WD & 8 & 13568 \\ \cline{1-6} 
\multirow{9}{*}{256} &  $64+12$            & CRC-Polar & 5G & 32 & 64 \\ \cdashline{2-2}\cline{3-6} 
                     & \multirow{2}{*}{64}  & PAC      & 5G & 16 & 48 \\ \cline{3-6} 
                     &                     & PAC      & 5G-WD & 32 & 1976 \\ \cline{2-6} 
                     &   $128+12$           & CRC-Polar & 5G & 16 & 371 \\ \cdashline{2-2}\cline{3-6}  
                     &  \multirow{2}{*}{128} & PAC      & 5G & 8 & 96 \\ \cline{3-6} 
                     &                     & PAC      & 5G-WD & 16 & 12096 \\  \cline{2-6} 
                     &  $192+12$            & CRC-Polar & 5G & 8 & 388 \\ \cdashline{2-2}\cline{3-6} 
                     & \multirow{2}{*}{192} & PAC      & 5G & 8 & 36416 \\ \cline{3-6} 
                     &                     & PAC      & 5G-WD & 8 & 36416 \\ \hline 
\end{tabular}
\end{table}


One way to form the information set $\I$ is to collect $\bigcup_{s=0}^{r-1} \S_{m-s} \subseteq\I$, which gives the dimension $\sum_{s=0}^{r-1} {m \choose s} \leq K$, and the rest (if needed) from $\S_{m-r}$ or equivalently from $\Z_{r}$, starting from level $|\lambda_f|=0$ while considering the reliability within the elements of the level. This approach results in excellent codes for small $m$. However, as we discussed the significant growth of $\Delta^{(m-s)}$ by $m$ in Section \ref{sec:pw_prop}, the reliability of elements at high levels, where $|\lambda_f|$ approaches $s(m-s)$, relative to the adjacent subset $\Z_{r+1}$ becomes very weak as $m$ increases. 

Alternatively, we propose to adjust a code constructed purely based on the reliability order of sub-channels to enhance its weight distribution. The adjustment is based on the idea discussed above for the first approach, dealing with two adjacent subsets $Z_r$ and $Z_{r-1}$, where $r=\max(\deg(f)),f\in\I$. Algorithm \ref{alg:enhanced5G} illustrates the details of this procedure based on the 5G reliability sequence $\Q$ and the contribution structure obtained in line 2. As line 8 demonstrates, we remove $\pi$ elements with degree $r$ and maximum $|\lambda_f|$ and minimum reliability with elements with degree $r-1$ and minimum $|\lambda_f|$ and minimum reliability. This order is given by flattening the contribution structure $\Z_s$ by function $FlatenZs(\cdot)$. 
Note that the two aforementioned approaches may give the same code.

\begin{figure}[ht]
    \centering
    \includegraphics[width=0.9\linewidth]{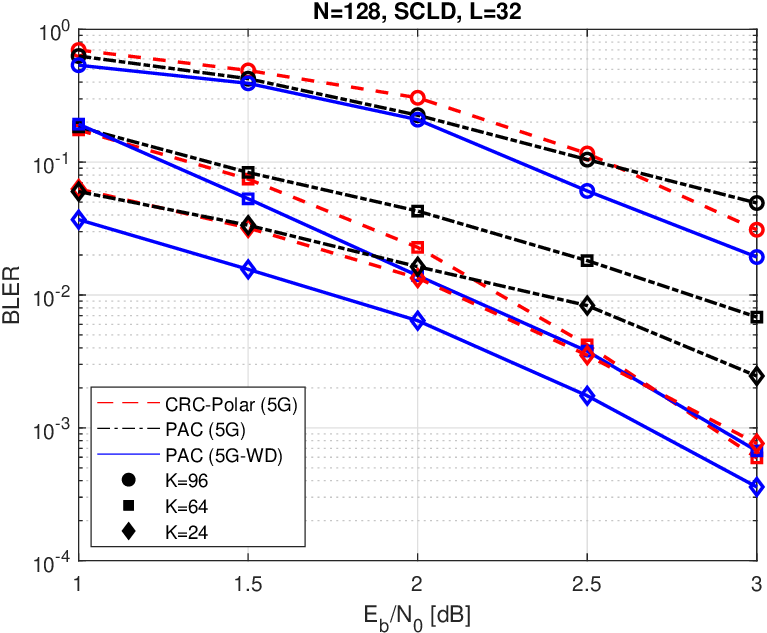}
    \caption{The performance comparison of the 5G reliability sequence and 5G-WD enhanced sequence at code length $N=128$.}
    \label{fig:BLER_128}
\end{figure}

\begin{figure}[ht]
    \centering
    \includegraphics[width=0.9\linewidth]{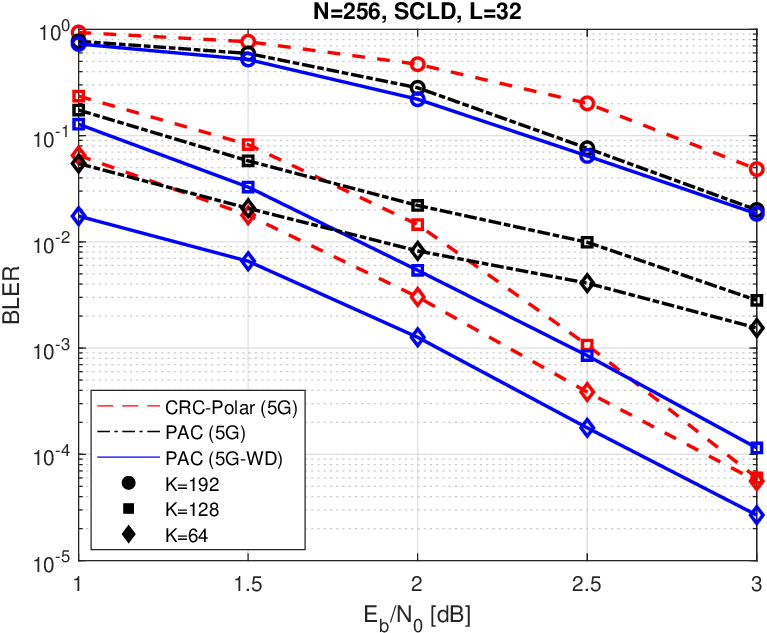}
    \caption{The performance comparison of the 5G reliability sequence and 5G-WD enhanced sequence at code length $N=256$.}
    \label{fig:BLER_256}
\end{figure}

Figs. \ref{fig:BLER_128} and \ref{fig:BLER_256} compare the error performance of the obtained rate-profile 5G-WD for various codes with block-length $N=128,256$ and $K/N=1/4,1/2,3/4$ using Algorithm \ref{alg:enhanced5G} for $\pi_{\max}=5$. As the results show, the obtained rate-profiles under PAC coding outperform other codes within the practical range of BLER ($10^{-2}-10^{-4}$). The $d_{\min}$ and $A_{d_{\min}}$ of the corresponding codes in Table \ref{tab:wd} indicates that although for some CRC-polar codes, $A_{d_{\min}}$ is smaller, the utilization of weak sub-channels for CRC bits weakens these codes, particularly at low rates. This reinforces that we need to balance the reliability and weight distribution of the codes.

Future work could explore alternative approaches that leverage weight contribution structures, such as the diagram shown in Fig. \ref{fig:monomials_deg2m6}. Ideally, the ordering of sub-channels should be designed to effectively incorporate both reliability and weight contribution. The selection of information bits based on such ordering should ensure that the resulting code exhibits favorable weight distribution and reliability. 

\begin{algorithm}[h]
\footnotesize

\KwIn{Sequence $\Q$, 
Parameters $N$, $K$, $m$, $\wm$, $\pi_{\max}$}
\KwOut{Rate-profile $\I$}

\SetKwFunction{CountOnes}{countOnes}
\SetKwFunction{BitReversed}{bitreversed}
\SetKwFunction{PW}{pw}
\SetKwFunction{MLLR}{mllr\_dega}
$\Q_N \gets [q \in \Q \mid q < \texttt{N}]$, $\Q_K \gets \Q_N[N-K:end]$\; 
$\Z \gets$ \FSub{$m$} \tcp*{Algorithm \ref{alg:groups}}
$\Z_{r}, \Z_{r-1} \gets  \Z[m-\log_2(\wm)], \Z[m-\log_2(\wm)-1]$\; 
\SetKwFunction{FSubF}{FlatenZs} 
$\S_{\wm}, \S_{2\wm} \gets$ \FSubF{$\Z_r,m,r$}, \FSubF{$\Z_{r-1},m,r-1$}\;
$\Q_{w_{\min}} \gets \S_{w_{\min}} \cap \Q_K$\;
$\overline{\Q}_{2w_{\min}} \gets \S_{2w_{\min}} \backslash Q_K$\;

$\pi \gets \min(\left(\min(\Q_{w_{\min}}, \overline{\Q}_{2w_{\min}}),\pi_{\max}\right)$\;
$\I \gets \left(\Q_K\backslash\Q_{w_{\min}}[1:\pi]\right)\cup\overline{Q}_{2w_{\min}}[1:\pi]$

\Return $\I$ \;

\SetKwFunction{FSubF}{FlatenZs} 
\SetKwProg{Fn}{function}{:}{} 
\Fn{\FSubF{$\Z_s,m,s$}}{
    \For{$d\gets 0$ \KwTo $s(m-s)$}{
        $\S \gets \S+$ Sorted $\Z_s[d]$ in DESC Rel. order\;
    } 
    \KwRet $\S$;
}
\caption{Enhancing weight distribution of 5G reliability sequence $\Q$}
\label{alg:enhanced5G}
\end{algorithm}

\section*{Acknowledgment}

We gratefully acknowledge Prof. Jinhong Yuan for supporting this work in part through the Australian Research Council (ARC) Discovery Project (DP220103596) and Linkage Project (LP200301482), both awarded to him.

\clearpage
\printbibliography

\end{document}